\documentclass[journal]{IEEEtran}
\IEEEoverridecommandlockouts

\usepackage{cite}
\usepackage{amsmath,amssymb,amsfonts}
\usepackage{algorithm}
\usepackage{algorithmic}
\usepackage{graphicx}
\usepackage{textcomp}
\usepackage{subcaption}
\usepackage{xcolor}
\usepackage{multirow}

\usepackage{graphicx}
\usepackage{epstopdf}

\title{\LARGE \bf
Creation of Synthetic Networked PMU Data:\\
A Generative Adversarial Network Approach
}

\author{ Xiangtian Zheng, Bin Wang, Dileep Kalathil and  Le Xie
\thanks{Authors are with the Department of Electrical and Computer Engineering at Texas A\&M University. Email: \texttt{\{ zxt0515,  binwang, dileep.kalathil, le.xie\} @tamu.edu}}
}

\begin{document}

\maketitle
\thispagestyle{empty}
\pagestyle{empty}

\begin{abstract}
This paper introduces a machine learning-based approach to synthetically creating realistic phasor measurement unit (PMU) data streams of multiple transient types. In contrast to the existing literature of transient simulation-based data generation methods, we propose a generative adversarial network (GAN) based approach to learning directly from the historical data and simultaneously reproduce multiple PMU data streams. The synthetic PMU data streams reflect meaningful dynamic characteristics which observe first principles such as Kirchhoff's laws. The efficacy of this approach is demonstrated by numerical studies on the IEEE 39-bus system. We validate the fidelity and flexibility of the synthetic data via statistical resemblance and modal analysis approaches. Finally we illustrate a practical application scenario for the usage of the synthetic PMU data, i.e. leverage the synthetic data to improve the performance of the event classification algorithms. 

\end{abstract}

\section{Introduction}
This paper aims at developing a new machine learning-based approach to synthetically creating massive data sets of PMUs that can be potentially used for data-intensive research purposes. Over the past decade, thousands of PMUs have been deployed in backbone transmission systems across North America and around the world. The tremendous amount of PMU measurements provide unique opportunities to improve the grid monitoring and control. Transient dynamic data captured by PMU are of particular value to the research community. However, there are only limited cases of transient real-world data recorded by PMU, and more importantly, these data are mostly confidential, e.g. as specified in the U.S. Critical Energy/Electric Infrastructure Information (CEII) that is protected by the Energy Act. On the other hand, a number of approaches using PMU data have been proposed in the recent years with promising results. This further increases the thirst for (transient) data. The lack of  availability of  the  transient data  becomes a critical bottleneck to data-driven research and development. To this end, creating realistic-looking data that are readily available to the research community seems to be an effective solution.

Research work on synthetic data creation in power systems comprises of two major thrusts: (i) acquiring data from simulation on realistic synthetic grids, and (ii) learning the characteristics with the training data set and then reproducing synthetic data that possess these learned features.  Majority of the efforts focus on creating a credible synthetic power network \cite{sg} - \cite{randomwalk}. With a synthetic power network, contingencies are then specified and the post-contingency responses of the system are simulated by computer simulation software such as PSS/E or PowerWorld. A lot of research efforts have been devoted to develop various synthetic network models, such as small-world \cite{smallworld} and random-walk model \cite{randomwalk}.   Two fundamental issues for the creation of synthetic power grids were addressed \cite{sg}, namely, geographic load and generator substation placement and assignment of transmission line electrical parameters. From a more practical perceptive, \cite{sg2} and \cite{sg3}  addressed the power flow convergence problem and dynamics in the  synthetic networks. Several large-scale synthetic network cases were released in \cite{ieee39}, which are freely available to the public.

As an alternative, there has been recent efforts of introducing machine learning concepts such as Generative Adversarial Network (GAN) for synthetic data generation. GAN can progressively learn the underlying characteristics of the limited-volume training data set to generate any large amount of credible synthetic data with realistic characteristics and rich diversity \cite{model-free} -\cite{GAN+stability}. GAN-based data generation approaches are simulation-free and also free of several common issues like inaccurate generator and load modeling, and low simulation efficiency. In recent years, several data-driven GAN-based applications have been proposed, such as renewable energy data generation \cite{model-free}, single PMU data generation \cite{sgsma} and dynamic security assessment \cite{GAN+DSA}\cite{GAN+stability}. 

Simulation-based data generated by the synthetic grids are commonly useful for the generic proof of concept, where any system model is acceptable as long as it is representative. However, motivated by actual needs, researchers may want to have a better understanding of a specific real system or implement novel applications in this system. In such   situations, only PMU data from this target system can help to do research or develop products. Without public access to the detailed real system model, simulation-based methods become restricted while GAN-based data-driven methods are appealing.

Ideally synthetic PMU data should possess two properties: \textit{fidelity} and \textit{flexibility}. Fidelity means that the synthetic data should have the same characteristics as real ones. Flexibility means that we can leverage the GAN model to controllably generate different classes of PMU data. However, those existing GAN-based approaches mentioned are not capable of creating PMU data streams from multiple locations in a network that possess meaningful dynamic characteristics and satisfy Kirchhoff's laws. To address these issues, this paper develops a GAN-based method to be able to synthesize meaningful networked PMU data streams, leveraging the analysis on the spatial and temporal correlation between multiple PMU data streams. Moreover, we illustrate a practical application scenario of the synthetic transient PMU data, i.e. data-driven event classification.

\subsection{Literature Review}
GAN was proposed in  \cite{gan}  which has now arguably became one of the most popular and successful deep generative models. While it drew increasing attention from multiple domains, GAN also faced the problems like vanishing gradients, mode collapse and training instability. To mitigate or even solve these problems, substantial theoretical research work was conducted, among which Wasserstein GAN \cite{wgan2} and Improved Wasserstein GAN \cite{im_wgan} are the representatives. To distinguish from the other extensions, we will call the original GAN as vanilla GAN in the rest of this paper.

Although GAN is a relatively new concept, it has achieved a great success in computer vision and a few other fields. For example, CycleGAN \cite{cyclegan} was proposed to transfer images from one domain to another, e.g. from real sceneries to Van Gogh style paintings. MIDINET \cite{midinet} aims at symbolic-domain music generation by a novel conditional mechanism and the trained model could be expanded to generate multiple realistic MIDI channels. SeqGAN \cite{seqgan} was proposed to respectively generate discrete and continuous time series data. Another paper \cite{cgantime} utilized conditional GAN to learn and simulate time series data, of which the conditions could be both categorical and continuous variables including different kinds of auxiliary information. DoppelGANger \cite{doppelganger} was proposed to model time series and mixed-type data through a hierarchy architecture, in which metadata and time series are generated separately and metadata are used to strongly influence time series generation.


However, there are only few efforts applying GAN to power systems. One representative is the renewable scenario generation \cite{model-free}, which generates realistic-looking wind and photovoltaic power profiles with rich diversities under various scenarios of interests. The first effort applying GAN to PMU data generation is presented in \cite{sgsma}, which learns the dynamics represented by the training dataset sampled from a single PMU and then produces a single synthetic PMU data stream. Another recent work \cite{GAN+DSA} used GAN to generate missing data of multiple PMUs in order to improve the performance of dynamic security assessment. However, it is not clear if that approach can be useful for more general applications. Moreover, GAN was applied to learn the spario-temporal correlations among historical local marginal price and predict the prices for the next hour \cite{GAN+lmp}.

\subsection{Main Contribution}
This paper aims at developing a new systematic approach to creating massive PMU data sets that can be readily accessible by the research community and useful for the subsequent data-driven methodologies. Main contributions are as follows: 

\begin{enumerate}
    \item \emph{Networked PMU Data Generation}: The proposed method can generate multiple realistic-looking networked PMU streams that respect the physical constraints like Kirchhoff's laws.
    \item \emph{High Efficiency}: Leveraging the temporal and spatial correlations among PMU streams, we show that we only need to synthesize $n_g$ PMUs by GAN where $n_g$ is the number of generators in the system which is  much smaller than the total number of PMUs in the system. This significantly improves the learning efficiency and reduces the computational burden.  
    \item \emph{Comprehensive Validation}: The efficacy and effectiveness of the synthetic PMU data are comprehensively validated from the perspective of statistical properties and modal analysis.
    \item \emph{Application Scenario Illustration}: We show that synthetic data generated can be used to improve the event classification accuracy. 

\end{enumerate}

The rest of this paper is organized as follows: Section II briefly reviews the basic mechanism of unconditioned vanilla GAN and conditional GAN. Section III executes the analysis on temporal and spatial correlations between voltage profiles at all buses and proposes the architecture of the networked PMU data generation. Section IV validates synthetic PMU data via the statistical resemblance, the modal analysis and the illustrated application scenario, i.e. event classification. Section V draws conclusions and states our future work. 
\section{Brief Review of GAN}
\subsection{Vanilla GAN}
GAN is a deep generative model that aims to generate realistic data. The fidelity of a GAN algorithm is measured by the difference between the distributions of the synthetic  data and real data. Therefore, the objective of GAN is to make the synthetic distribution close enough to the real one.

\begin{figure}[htpb!]
	\center
	\includegraphics[width=0.48\textwidth]{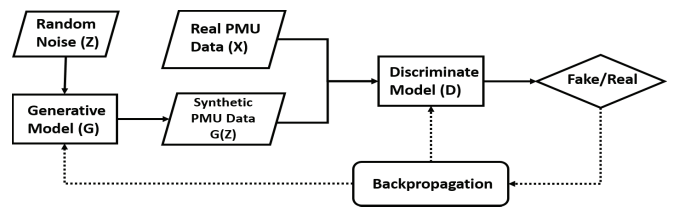}
	\caption{Diagram of GAN}
\end{figure}

Fig. 1 shows the architecture of vanilla GAN \cite{gan} with the following notations: noise data $z$ follow a known distribution such as a multivariate Gaussian distribution $\mathop{\mathbb{P}}_Z$, while the real data follow a certain but unknown distribution $\mathop{\mathbb{P}}_X$.

The generative model (generator) $G$ is trained to be a function mapping from $\mathop{\mathbb{P}}_Z$ to $\mathop{\mathbb{P}}_X$. The input data $z$ are easily sampled from the pre-specified distribution $\mathop{\mathbb{P}}_Z$. The distribution of synthetic data generated by $G$ model is denoted as $\mathop{\mathbb{P}}_g$. Given a batch of noise data as input, a well-trained generative model should generate a batch of diverse realistic-looking data whose distribution $\mathop{\mathbb{P}}_g$ is supposed to close enough to $\mathop{\mathbb{P}}_X$. 

The discriminate model (discriminator) $D$ takes input data sampled from either the real data set or the synthetic data set. Its output, a scalar ranging from 0 to 1, indicates the likelihood that the input data belongs to the real data set. The goal of discriminate model $D$ is to correctly distinguish $\mathop{\mathbb{P}}_X$ from $\mathop{\mathbb{P}}_g$ by maximizing the difference between the output scalars of real and synthetic samples.

With the above descriptive definition of $G$ and $D$ models, the two cost functions in vanilla GAN are formulated as follows:
\begin{equation}
J_D = -\mathop{\mathbb{E}}\limits_{x\sim\mathop{\mathbb{P}}_X}\log(D(x))-\mathop{\mathbb{E}}\limits_{z\sim\mathop{\mathbb{P}}_Z} \log(1-D(G(z))) 
\end{equation}
\begin{equation}
J_G = -\mathop{\mathbb{E}}\limits_{z\sim\mathop{\mathbb{P}}_Z} \log(D(G(z))) 
\end{equation}

The above two cost functions can be combined into a single objective function $J$:
\begin{equation}
\min_{G}\max_{D}J=\mathop{\mathbb{E}}\limits_{x\sim\mathop{\mathbb{P}}_X}\log(D(x))+\mathop{\mathbb{E}}\limits_{z\sim\mathop{\mathbb{P}}_Z} \log(1-D(G(z))) 
\end{equation}

To optimize the objective function in (3), the common training procedure of this minimax game is listed below:

(a) Initialize parameters in both $G$ and $D$ models.

(b) Randomly pick $k$ samples from the real data set and another $k$ samples from the pre-specified noise distribution (a uniform distribution is used in this paper), where $k$ is the pre-specified size of the minibatch.

(c) Update $D$ model for $k_D$ times with backpropagation according to the derivative of the objective function w.r.t. $\theta_D$, where $\theta_D$ is the parameter of the discriminator and $k_D$ is a constant integer.

(d) Update $G$ model for only once with backpropagation according to the derivative of the objective function w.r.t. $\theta_G$, where $\theta_{G}$ is the parameter of the generator.

(e) Iterate steps (b) to (d) until the number of steps reaches the preset number or losses of both models converge.

\subsection{Conditional GAN}
 Different from the  vanilla GAN, conditional GAN can control the specific types of the samples of our interests, which matches the expectation of flexibility. Conditional generation is done by incorporating more information into the training procedure of GANs such that the generated samples conforming to same properties as certain class of training samples. Formally, the objective function $J$ can be written as:
 
 \begin{equation}
\min_{G}\max_{D} \mathop{\mathbb{E}}\limits_{x\sim\mathop{\mathbb{P}}_X}\log(D(x|y))+\mathop{\mathbb{E}}\limits_{z\sim\mathop{\mathbb{P}}_Z} \log(1-D(G(z|y))) 
\end{equation}
where $y$ is the label that encodes different types.

Class labels are assigned based on the user-defined classification metrics, which are the representation of the events represented by the given samples, such as disturbance types, oscillation location, etc. Its training procedure is nearly same as that of unconditioned vanilla GAN, except that conditional GAN's input has an additional label vector. Once training loss converges, conditional GAN should be able to learn the conditional distribution and generate samples based on any meaningful classification metrics. 

\subsection{Conditional GAN with Gradient Penalty Regularization}
Many modified objective functions for GAN have been proposed  to overcome common issues, such as mode collapse, gradient vanishing and unstable training. To achieve a better convergence performance, we choose the zero-centered gradient penalty method \cite{which_gan} as the regularization of the model $D$. To alleviate the mode collapse, we choose the mode seeking penalty\cite{msgan} as the regularization of the model $G$.

Formally the regularization for $G$ and $D$ model can be respectively written as:

\begin{equation}
    \Omega_G = \beta_G\mathop{\mathbb{E}}\limits_{z\sim\mathop{\mathbb{P}}_Z} 1/||\bigtriangledown D(G(z|y))||^2
    \label{eq:regularization_G}
\end{equation}
\begin{align}
    \Omega_D & = \beta_{DD}\mathop{\mathbb{E}}\limits_{x\sim\mathop{\mathbb{P}}_X} ||\bigtriangledown D(x|y)||^2 \notag\\
    & +\beta_{DG} \mathop{\mathbb{E}}\limits_{z\sim\mathop{\mathbb{P}}_Z}||\bigtriangledown D(G(z|y))||^2
\end{align}
where $\beta_G$, $\beta_{DD}$ and $\beta_{DG}$ are user-defined weights.

In conclusion, the algorithm utilized in this paper is described  below, where $T_{\max}$ is a user-defined maximum number of iteration rounds, $k_D$ is the number of times of consecutively updating $D$ model and $k$ is the minibatch size. We will introduce the implementation details in the Section IV.

\begin{algorithm}[htpb!]
 	\caption{Conditional GAN with Gradient Penalty Regularization}
 	\label{algo:marl}
 	\begin{algorithmic}
 		\STATE Initialize $\theta_D$ and $\theta_G$ 
 		\FOR {$T = 1$ to $T_{max}$} 
 		\FOR {$j=1$ to $k_D$}
 		\STATE Sample real data $\{x_i\}_{i=1}^k$ and labels $\{y_i\}_{i=1}^k$
 		\STATE Sample latent variables $\{z_i\}_{i=1}^k$
 		\STATE $\theta_D\leftarrow\theta_D+{\rm AdamProp}(J_D-\Omega_D)$
 		\ENDFOR
 		\STATE Sample latent variables $\{z_i\}_{i=1}^k$
 		\STATE $\theta_G\leftarrow\theta_G-{\rm AdamProp}(J_G+\Omega_G)$
 		\ENDFOR
 	\end{algorithmic} 
 \end{algorithm}

\section{GAN-based Networked PMU Data Generation Approach}
\subsection{Problem Statement}
Consider a power system with $n$ buses and $m$ branches (including the lines and transformers). We denote the voltage at bus $i$ by $V_{i}, i=1,...,n$. Since there are $m$ branches, we assume that there are $2m$ current measurements (one on each end), and are denoted by $I_k, k=1,...,2m$. We assume that the bus admittance matrix is known. However, we don't assume the knowledge of the model of the generator dynamics which is typically very difficult to obtain. 

Consider a set of labeled historical data obtained from the real world PMU measurements, where the labels are user-defined. For each bus $i$, historical voltage data is denoted by $V_{i}(t)$, and for each end of a branch $k$, historical current data is denoted by $I_{k}(t)$ for time index $t=1, \ldots, T$. Our goal is to develop a GAN-based algorithm for generating the synthetic PMU data $\tilde{V}_{i}(t), \tilde{I}_{k}(t)$ using the labeled historical data as the training data in such way that given a label the synthetic data generated exhibits all the relevant statistical properties exhibited by the same class of historical data. Moreover, the synthetic data should satisfy the physical constraints imposed by the Kirchhoff's voltage and current laws at each time $t$.

The synthetic PMU data should be meaningful in such a way that the data generated at all the buses and the branches should satisfy the spatial and temporal correlation property of the real data. This is significantly different from the standard application of GAN for image generation \cite{gan} or even wind and solar data generation in power system \cite{model-free} which can be more appropriately  be modelled as statistical phenomenons. However, in the PMU data generation, the spatial correlation is through the Kirchhoff's voltage and current laws which have to be satisfied at each time deterministically. The temporal correlation in the data is induced by the generator dynamics which also has unique physical characteristics. So, a straightforward application of GAN for generating data at each bus \cite{sgsma} may not yield realistic network level synthetic PMU data. In the following, we propose a clever way for doing the same. 

\subsection{Power Systems Dynamics and Reduced Model}
Before developing the algorithm, we first give a brief description about the power systems dynamics. The intuition from this domain knowledge is then used to precisely characterize the spatial and temporal correlation, which are then exploited in developing the network level GAN algorithm. 

The dynamic model of a power system contains two parts: differential equations and algebraic equations \cite{psds}. Differential equations characterize the temporal dependence through the  the dynamics of inner states of generators and the algebraic equations characterize the spatial dependence though the  Kirchhoff's laws. Assume there are $n_g$ generator buses in the system and all loads are represented by the constant impedance model. Let $x_{i}$ denotes the internal state of generator $i$, usually including rotor angle $\delta_i$, angular speed $\omega_i$ and other transient inner variables. Then the dynamics of the $i$th generator is given by
\begin{equation}
\label{eq:dynamics}
    \dot{x}_{i} = f_{i}(x_{i}, V_{i}, I_{dqi}),
\end{equation}
where $I_{dqi}$ is the dq-axis currents. 

The algebraic equations have two parts, stator algebraic equation and network algebraic equation. The stator algebraic equation of the $i$th generator is given by
\begin{align}
0=&{|V_i|}e^{j\theta_i}+(R_{si}+jX_{di}')(I_{di}+jI_{qi})e^{j(\delta_i-\pi/2)} \nonumber \\
\label{eq:stator}
&-[E_{di}'+(X_{qi}'-X_{di}')I_{qi}+jE_{qi}']e^{j(\delta_i-\pi/2)}
\end{align}
where $R_{si}'$, $X_{di}'$ and $X_{qi}'$ are constant resistance and impedance parameters. $|V_i|$ and $\theta_i$ are the magnitude and angle of the terminal voltage $V_i$. $E_{di}'$ and $E_{qi}'$ are inner voltage state variables included in $x_i$. This equation can be succinctly represented as

\begin{equation}
\label{eq:stator2}
I_{dqi}=h_i(x_i, {V}_i). 
\end{equation}

Because of the assumption of the constant impedance load, load can be incorporated into the admittance matrix $Y$, making the equivalent current injection at load buses equal to 0. Then the network equation is given by 
\begin{equation}
\label{eq:voltage}
\left[
\begin{matrix}
I_1^{\text{inj}}\\
...\\
I_{n_g}^{\text{inj}}\\
0\\
...\\
0\\
\end{matrix}
\right]_{n\times 1}=\left[
\begin{matrix}
Y
\end{matrix}
\right]_{n\times n}\left[
\begin{matrix}
{V}_1\\
...\\
{V}_n\\
\end{matrix}\right]_{n\times 1}
\end{equation}
where $Y$ is the admittance matrix that is already incorporated with the constant impedance load and $I_i^{\text{inj}}$ is the injected current from generator $i$. 

One can show that, using Kron reduction \cite{psds}, the network with $n$ buses can be reduced to a smaller system with just $n_g$ generator internal buses. We will use this result in developing the network GAN algorithm. 

\subsection{Networked PMU Data Generation}
Using the reduced model,  it is sufficient to generate  synthetic PMU data only for the generator buses. The PMU data for the other buses can be generated from this data by exploiting the temporal and spatial correlation specified by the power system model. Since the bus admittance matrix $Y$ is invertible for any operable system, \eqref{eq:voltage} implies that only $n$ of total $n+n_g$ variables are linearly independent. Suppose the voltage at all generator buses are available, then the voltages at all the other buses can be computed using \eqref{eq:voltage} with the known admittance matrix. 

However it is not sufficient to generate independent PMU data for each generator bus separately using the method in \cite{sgsma} because of the temporal correlation between data at the generator buses. To see this, assume that the internal state of $i$th generator at time $t$, $x_{i}(t)$, is available for all $1 \leq i \leq n_{g}$. Then $V_{i}(t)$ and $I_{dqi}(t)$ can be computed using \eqref{eq:stator2} and \eqref{eq:voltage}. Note that this computation requires the voltage at all generator buses, not just the voltage at bus $i$. Given $V_{i}(t)$ and $I_{dqi}(t)$, $x_{i}(t+1)$ can be computed using \eqref{eq:dynamics}. This procedure can be repeated to generate the PMU data for all the generator buses, and by the argument above, for all the buses in the network. Due to the temporal correlation induced in the intermediate step, voltage profiles at the generator buses have to be generated simultaneously by one single GAN model. In contrast, the naive idea of training multiple GANs to learn individual profiles and then simply aggregate individual synthetic profiles will not work because it does not consider the temporal correlation.

\begin{figure}[htpb!]
	\center
	\includegraphics[width=0.48\textwidth]{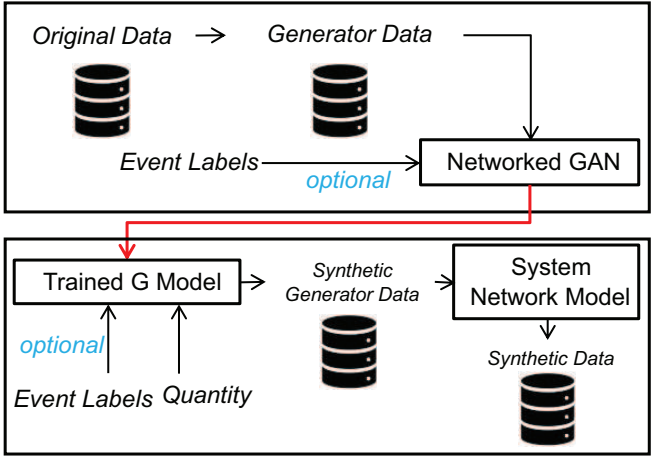}
	\caption{Diagram of networked PMU data generation. The label vector input of $G$ model is optional, depending on whether the real dataset is labeled.}
	\label{fig:networked_PMU}
\end{figure}

We now give a  summary of the networked PMU data generation algorithm, along with a diagram in Fig. \ref{fig:networked_PMU}. We will give a detailed description of step 1 to 3 in Section IV.

\begin{enumerate}
    \item Pre-processing: Normalize voltage profiles at all generator buses and then truncate and downsample the data. 
    \item GAN training: Use the labeled historical voltage profiles at all generator buses simultaneously to train a GAN, following the process shown in Algorithm \ref{algo:marl}. Output is the realistic looking voltage profiles at all generator buses.
    \item Post-processing: Filter out the high frequency noise in the generated voltage profiles. Renormalize the voltage profile to match the real data.
    \item Recovering: Calculate all voltage profiles at non-generator buses and current profiles using \eqref{eq:voltage} to reconstruct a complete synthetic event recorded by the PMUs over the whole system. 
\end{enumerate}

\section{Transient Networked PMU Data Generation}

In this section, we first give a detailed description of power system simulation and algorithm implementation. We then analyze its performance using several metrics. In particular, we show that the voltage profiles generated using our approach are visually indistinguishable from the real data. They also have similar statistical properties. Moreover, we use Prony analysis \cite{mam} to show that the system dynamics reflected by the synthetic data has really similar characteristics to that of the real data. Further, we take a binary event classification task as the application scenario to illustrate one possible usage of the synthetic transient PMU data, verifying that additional synthetic training data can improve the classification accuracy.

\subsection{Training Data Generation and Processing}
\begin{figure}[htpb!]
	\center
	\includegraphics[width=0.9\columnwidth]{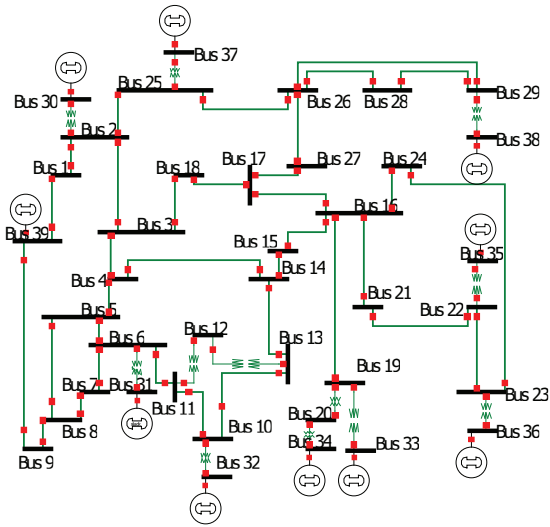}
	\caption{Diagram of IEEE 39-bus system\cite{ieee39}}
	\label{fig:IEEE39}
\end{figure}
We use the IEEE 39-bus system shown in Fig. \ref{fig:IEEE39} for training and testing the proposed algorithm. We take the simulation data as the real data, due to the unavailability of real-world PMU data.

As mentioned in the introduction, we are particularly interested in transient data, meaning post-fault data that reflect the system dynamics. To reflect the flexibility of the generative model, we would like to controllably generate transient data caused by different faults. Therefore 3-phase self-clearing faults and line tripping cases are simulated with random parameters. The post-fault voltage profiles at all generator buses together with their event labels are recorded as the training and test dataset for the GAN. Simulation settings are listed in Table \ref{table:test_system}. Note that for each event type, only $50$ samples will be used as the training dataset while the rest will be the test dataset.

\begin{table}
	\begin{center}
		\caption{Simulation configuration}
		\begin{tabular}{ccc}
			\hline
			Setting & Bus fault & Line tripping \\ \hline
			Fault duration &  0-0.1s & 0-0.1s\\
			Fault location &  random bus & random line\\
			Sampling frequency &  120 Hz & 120 Hz\\
			Time window & 7 s & 7 s\\ 
			Number of events & 1000 & 1000 \\
			Event label      & 0     & 1\\\hline
		\end{tabular}
		\label{table:test_system}
	\end{center}
\end{table}
\begin{table}
	\begin{center}
		\caption{Architecture of $G$ and $D$ Models}
		\begin{tabular}{ccc}
			\hline
			 Layer  & $G$ model & $D$ model\\ \hline
			 Input  & 32                    & 20$\times$20$\times$20\\          
			 Layer1 & Dense 128$\times$5$\times$5   & Conv64 2$\times$2 \\
			 Layer2 & De-Conv64 2$\times$2   & Conv128 2$\times$2 \\
			 Layer3 & De-Conv20 2$\times$2    & Dense 1 \\\hline
        \label{table:GAN_architecture}
		\end{tabular}
	\end{center}
\end{table}
To efficiently train the GAN, we truncate and downsample the raw data, while ensuring that the dynamic settling process is unchanged. Therefore both downsampling and truncation rate are determined by the settling speed of profiles in the real data set. Raw data are downsampled to 60 samples per second. We then truncate the first 400 points of each profile and normalize them to the range from 0 to 1. In the final step, we transform each 1-D profile into a 20-by-20 matrix and then concatenate them along the third dimension to match the input size of the generator model $G$.

After training, raw synthetic PMU data generated by GAN need post-processing: upsampling, renormalization and filtering. Upsampling is to make the sampling frequency of synthetic profiles 120 samples per second, the same as the real data. Renormalization is to convert the unit of synthetic data back to the normal scale. Filtering is to smooth out the fictitious high-frequency components by a low-pass filter. This is required because the learning algorithm by itself seems unable to do this even with larger network and careful tuning of the hyperparameters.   

\subsection{GAN Model Architecture and Training Procedure}
The generator $G$ and the discriminator $D$  are achieved by deep convolutional neural networks, inspired by DCGAN \cite{dcgan}. $G$ model contains 1 fully-connected layer and 2 de-convolutional layers with the stride size of 2$\times$2. $D$ model contains 2 convolutional layers with the stride size of 2$\times$2 and 1 fully-connected layer. Table \ref{table:GAN_architecture} lists the detailed architectures of $G$ and $D$ models adopted in the experiment, where the number after dense layers means the output size while those after convolutional or deconvolutional layers respectively mean the output dimension and stride size. All hidden layers in the $G$ model are equipped with ReLU activation function while leaky ReLU activation is applied to the hidden layers in $D$ model. Besides, batchnormalization technique is applied to all convolutional and deconvolutional layers in both models.

\begin{figure}[htpb!]
	\center
	\includegraphics[width=\columnwidth]{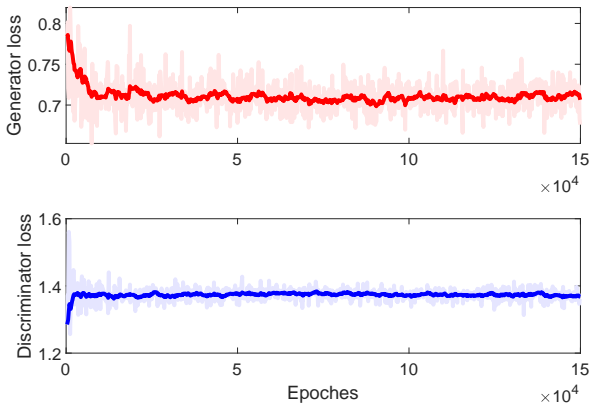}
	\caption{Convergence process of $G$ and $D$ loss functions}
  \label{fig:loss}
\end{figure}

Algorithm \ref{algo:marl} in  Section II  shows the training algorithm for GAN. In this experiment, we fix $T_{\max}=150000$ and $\beta_D=1$. Mini-batch size $N$ for each iteration of training is set to be $32$. All parameters in layers are initialized with the normal distribution. To achieve stable and quick convergence, GAN model is updated by AdamProp optimizer with a learning rate of $10^{-4}$. The training dataset consists of only $50$ samples for each type of event, which are part of $1000$ samples generated in the Section IV-A. The loss functions of both $D$ and $G$ models converge after about 20000 iterations as shown in Fig. \ref{fig:loss}. After training, we use the well-trained $G$ model to generate 1600 synthetic samples for each event type, for the following fidelity validation.

\subsection{Fidelity Validation: Statistical Similarity}
\begin{figure*}[htpb!]
	\begin{subfigure}{0.48\textwidth}
		\centering
		\includegraphics[width=1\textwidth]{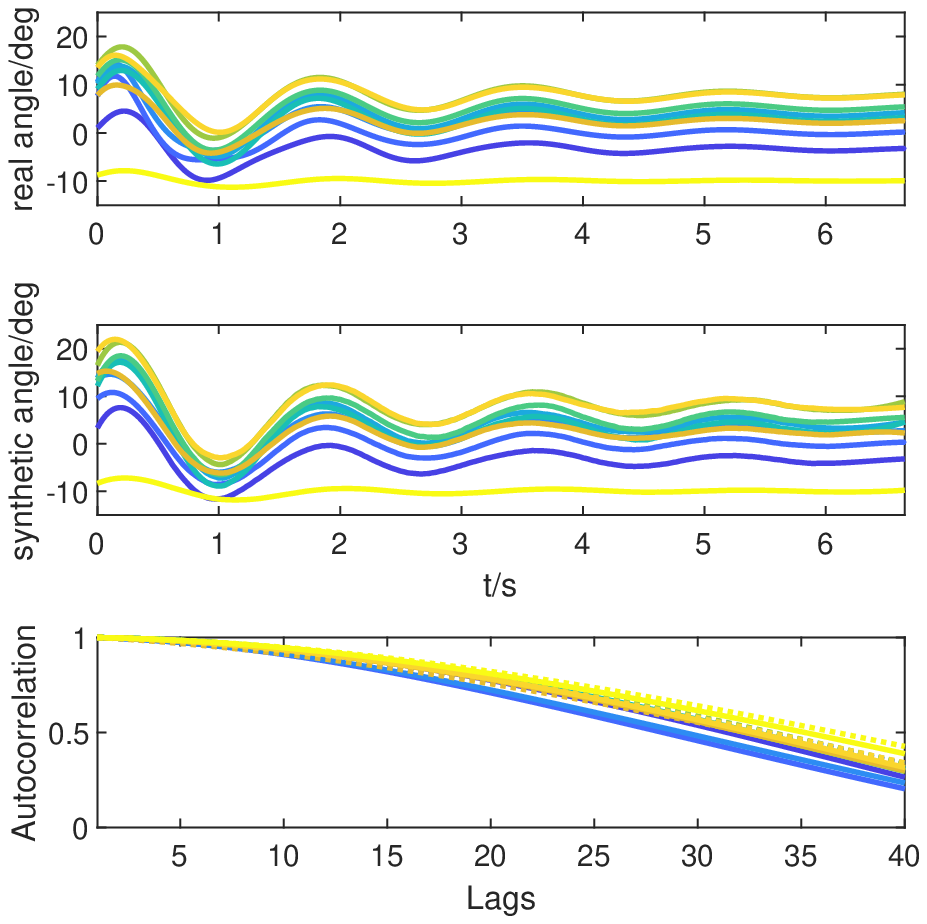}
		\caption{Bus fault case}
	\end{subfigure}
	\begin{subfigure}{0.48\textwidth}
		\centering
		\includegraphics[width=1\textwidth]{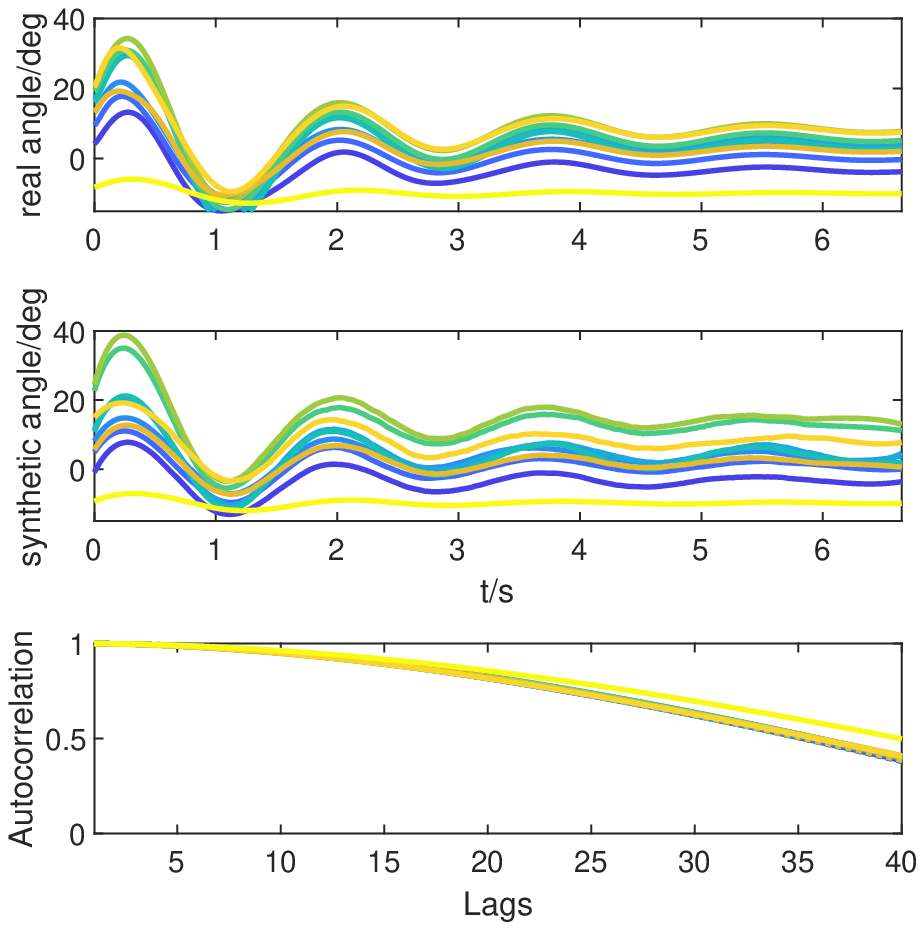}
		\caption{Line tripping case}
	\end{subfigure}
	\caption{Comparison between real and synthetic post-fault voltage angle profiles and their autocorrelation curves in the IEEE 39-bus system. In each subfigure, the top and middle rows respectively show the real and synthetic voltage angle profiles at all generator buses, while the bottom row compare their autocorrelation curves where solid lines mean real data and dotted lines mean synthetic data.}
	\label{fig:auto}
\end{figure*}

Firstly we validate that GAN models can generate transient PMU data that capture the inherent temporal correlation. For the visual comparison, top and middle rows in Fig. \ref{fig:auto} respectively illustrate the real and synthetic post-fault voltage angle profiles at all generator buses. It is clear from the visual comparison that: (i) real and synthetic profiles have similar settling patterns; (ii) the ranges of real and synthetic profiles are nearly the same.

We also test the statistical similarity using the metric of autocorrelation coefficient. This approach is commonly used to analyze the statistical correlations, e.g.  \cite{model-free} verified the realistic statistical properties of synthetic renewable energy data. In this experiment, we set the maximum lag as $40$ time steps, equal to 1/3 second. The bottom row in Fig. \ref{fig:auto} shows the comparison between the autocorrelation coefficient curves of real and synthetic profiles, illustrating the underlying temporal correlation. Real and synthetic autocorrelation coefficient curves approximately overlap with each other, qualitatively verifying the statistical resemblance. 

\subsection{Fidelity Validation: Modal Analysis}
Further we use modal analysis to quantitatively show that synthetic PMU data possess realistic dynamic characteristics of power systems, i.e. modal properties. Specifically, we use Prony analysis \cite{mam}, a classical ring-down analysis method, to analyze the synthetic data and extract important modal properties including oscillation frequency and damping. 

Since modes are the fingerprint of a given linear system, we can compare the modes of the real data and synthetic data for validating the fidelity of the synthetic data. Here we select the voltage angle measurements for validation. Details of the validation test are summarized as below:
\begin{enumerate}
    \item Modal Characteristics Estimation:  Calculate the oscillation frequency ($\omega_{i}$), damping coefficient ($\sigma_{i}$), amplitude ($a_{i}$), and phase ($\theta_{i}$) of each mode $i$ via Prony analysis for real and synthetic voltage profiles.
     \item Dominant Modes Selection: Only the modes with amplitude greater than a threshold are selected as the dominant mode. The threshold is fixed as the 10\% of the maximum amplitude.  
    \item Validation: For each synthetic profile, modal frequencies and damping ratios of the dominant nodes are compared with those of all real profiles at the same generator bus. We say that the synthetic profile passes the test if all dominant modes of the tested synthetic profile appear in the real data. We declare a pair of modes as the same if their relative error is less than $5\%$. 
\end{enumerate}

\begin{figure}[htpb!]
	\center
	\includegraphics[width=\columnwidth]{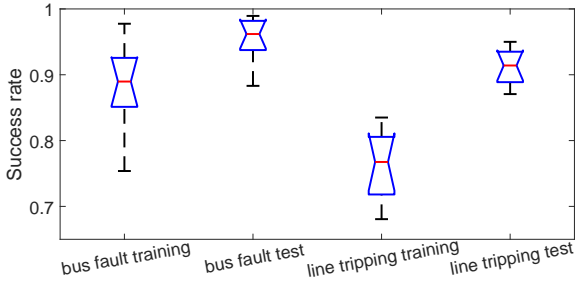}
	\caption{Success rate of the synthetic data of bus fault and line tripping events with training and test dataset being the benchmark}
  \label{fig:success_rate}
\end{figure}

We separately test the synthetic data of both event types, with training and test dataset generated in the Section IV-A being the benchmark respectively. Fig. \ref{fig:success_rate} shows the statistical results of single PMU's success rate. From these observations, we can conclude that: 
\begin{enumerate}
    \item \textit{Fidelity}: High success rate indicates that the synthetic data approximately capture the dynamic characteristics of the real data;
    \item \textit{Diversity}: The difference of the success rate between training and test benchmark indicates that the GAN model does not simply memorize the training data but also it creates new and meaningful modes that exist in the test dataset;
    \item \textit{Flexibility}: High success rates of both event types indicate that the GAN model can controllably synthesize certain realistic scenarios of interests.
\end{enumerate}

\subsection{Application Scenario Illustration: Data Enrichment for Event Classification}
\begin{figure}[htpb!]
	\center
	\includegraphics[width=\columnwidth]{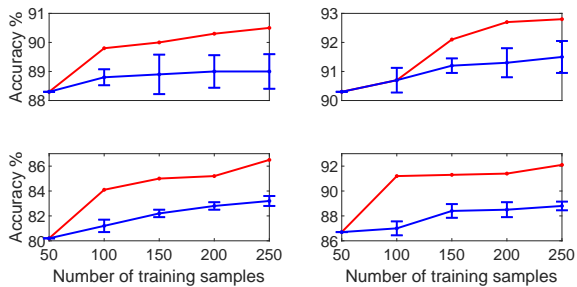}
	\caption{Event classification accuracy using real (red) and hybrid (blue) training data. Top left: SVM+DWT; top right: AdaBoost+DWT; bottom left: SVM+PCA; bottom right: AdaBoost+PCA.}
  \label{fig:event_classification}
\end{figure} 

The performance of data-driven methods is significantly determined by the volume and quality of the training dataset \cite{data_science}. However, training data shortage is a common dilemma in practice. In such situation, leveraging the GAN-based generation methods to enrich the training dataset may help to mitigate the data shortage. In this subsection, we will illustrate an application scenario to show that the GAN-based PMU data generation method can improve the performance of the subsequent data-driven method.

Specifically, we take binary event classification as the application scenario. Event classification tasks can be separated into two steps: feature extraction and feature classification. Considering the possible impacts brought by choosing different approaches, we select two common methods respectively for both steps. Traditional pattern recognition methods such as Discrete Wavelet Transform (DWT) \cite{svm-event} and Principle Component Analysis (PCA)\cite{pca_pmu} are selected to extract the high-level features. DWT uses the mother wavelet, which is a set of basic functions decomposing the data into several resolution levels. The coefficients, which carry the detailed and approximate information of the data, can be used as features for pattern recognition. PCA projects the data onto the principle subspace such that the variance of the data is maximized. The dynamics of the data can be analyzed by transferring the data into a combination of principle component. Traditional classifier models such as Support Vector Machine (SVM) \cite{svm-event} and Adaptive Boosting (AdaBoost) are used to implement the classification task. SVM enables mapping the original feature into a higher dimensional space so that the decision boundary can be identified. With decision trees as week learners, AdaBoost gathers the information about the relative  hardness of each training sample at each stage and then feeds them into the tree growing algorithm such that trees at later stage tend to focus on harder-to-classify samples. Finally AdaBoost combines the output of trained weak learners into a weighted sum to make a final decision. 

In this experiment, training dataset contains two events i.e. bus fault and line tripping. To verify the impacts brought by the volume and quality of the training data, we have real and hybrid training dataset, of which the size varies from 50 to 250. Hybrid training dataset consists of 50 real samples and additional synthetic samples that are generated by the GAN model trained on these real ones. For each training dataset, we respectively utilize DWT and PCA to extract the corresponding features, and then use SVM and AdaBoost to implement the classification. The trained models are examined by the test dataset generated in the Section IV-A. Fig. \ref{fig:event_classification} shows the comparison of the test accuracy of classification models trained by different training dataset. According to these observations, we have following conclusions:
\begin{enumerate}
    \item Different combinations of the feature extraction methods and classifier models have different impacts on the classification accuracy and its sensitivity to the volume of the training data;
    \item The shortage of the real training data always restricts the classification accuracy, no matter using which combination of the feature extraction and classifier models;
    \item Trained on the limited real data, GAN can reproduce additional realistic synthetic data that can mitigate the training data shortage problem and improve the classification accuracy.
\end{enumerate}

\section{Conclusion and Future Work}
In this paper, we proposed a Generative Adversarial  Network (GAN) based networked PMU data generation approach for synthesizing realistic transient PMU data in a network. Through a model reduction approach we significantly reduce the  volume of synthetic data directly generated by GAN. At the same time, our approach ensures the spatial correlation required by the Kirchhoff's laws and temporal correlation induced by the underlying generator dynamics. We validated the synthetic data in terms of statistical characteristics and modal properties. According to the validation results, synthetic PMU data have fairly strong fidelity and flexibility. Through the application scenario illustration, it is believable that the synthetic PMU data generated by the proposed approach can be useful for subsequent data-driven methods.

In our future work, we will improve the performance of the algorithm such that the synthetic data will be indistinguishable from the real ones by any advanced analytic method, e.g. existing power system dynamic analysis methods. We will also try this approach in the cases of systems with time-varying dominant modes and  will consider a more general case with non-constant-impedance loads. Both are challenging problems which require more analytical tools than used in this paper.



\bibliographystyle{plain}

\begin{thebibliography}{99}

\bibitem{sg}
K. M. Gegner, A. B. Birchfield, T. Xu, K. S. Shetye, and T. J. Overbye, "A methodology for the creation of geographically realistic synthetic power flow models," in Proc. IEEE Power Energy Conf. Illinois, 2016, pp. 1–6.
\bibitem{sg2}
A. B. Birchfield, T. Xu and T. J. Overbye, "Power Flow Convergence and Reactive Power Planning in the Creation of Large Synthetic Grids," in IEEE Transactions on Power Systems, vol. 33, no. 6, pp. 6667-6674, Nov. 2018.
\bibitem{sg3}
T. Xu, A. B. Birchfield and T. J. Overbye, "Modeling, Tuning, and Validating System Dynamics in Synthetic Electric Grids," in IEEE Transactions on Power Systems, vol. 33, no. 6, pp. 6501-6509, Nov. 2018.
\bibitem{smallworld}
D. J. Watts and S. H. Strogatz, "Collective dynamics of 'Small-World' networks," Nature, vol. 393, no.6684, pp. 393–440, 1998.
\bibitem{randomwalk}
Z. Wang, R. J. Thomas and A. Scaglione, "Generating Random Topology Power Grids," Proceedings of the 41st Annual Hawaii International Conference on System Sciences (HICSS 2008), Waikoloa, HI, 2008, pp. 183-183.
\bibitem{ieee39}
"Electric Grid Test Case Repository," Texas A\&M University College of Engineering. [Online]. Available: https://electricgrids.engr.tamu.edu/electric-grid-test-cases/.
\bibitem{model-free}
Y. Chen, Y. Wang, D. Kirschen and B. Zhang, "Model-Free Renewable Scenario Generation Using Generative Adversarial Networks," in IEEE Trans. on Power Systems, vol. 33, no. 3, pp. 3265-3275, May 2018.
\bibitem{sgsma}
X. Zheng, B. Wang and L. Xie, "Synthetic Dynamic PMU Data Generation: A Generative Adversarial Network Approach," 2019 International Conference on Smart Grid Synchronized Measurements and Analytics (SGSMA), College Station, TX, USA, 2019, pp. 1-6.
\bibitem{GAN+DSA}
C. Ren and Y. Xu, "A Fully Data-Driven Method based on Generative Adversarial Networks for Power System Dynamic Security Assessment with Missing Data," in IEEE Trans. on Power Systems, 2019.

\bibitem{GAN+stability}
X. Cao, G. Raman, G. Raman, and J. Peng. "Generative Adversarial Networks for Real-time Stability of Inverter-based Systems." arXiv preprint arXiv:1901.05114 (2019).

\bibitem{GAN+lmp}
Z. Zhang and M. Wu. "Predicting Real-Time Locational Marginal Prices: A GAN-Based Video Prediction Approach." arXiv preprint arXiv:2003.09527 (2020).
\bibitem{svm-event}
S. Ekici. "Classification of power system disturbances using support vector machines." Expert Systems with Applications. vol. 36. pp. 9859-9868, 2009. 
\bibitem{gan}
I. Goodfellow, J. Pouget-Abadie, M. Mirza, B. Xu, D. Warde-Farley, S. Ozair, A. Courville, and Y. Bengio, "Generative adversarial nets," Advances in Neural Information Processing Systems, 2014, pp. 2672–2680.
\bibitem{wgan}
M. Arjovsky, S. Chintala, and L. Bottou. "Wasserstein gan." arXiv preprint arXiv:1701.07875 (2017).
\bibitem{wgan2}
M. Arjovsky, S. Chintala, and L. Bottou. "Wasserstein generative adversarial networks." In International conference on machine learning, pp. 214-223. 2017.
\bibitem{im_wgan}
I. Gulrajani, F. Ahmed, M. Arjovsky, V. Dumoulin, and A. Courville. "Improved training of wasserstein gans." In Advances in neural information processing systems, pp. 5767-5777. 2017.
\bibitem{which_gan}
L. Mescheder, G. Andreas, and N. Sebastian. "Which training methods for GANs do actually converge?." arXiv preprint arXiv:1801.04406 (2018).
\bibitem{msgan}
Q. Mao, H. Lee, H. Tseng, S. Ma, and M. Yang. "Mode seeking generative adversarial networks for diverse image synthesis." In Proceedings of the IEEE Conference on Computer Vision and Pattern Recognition, pp. 1429-1437. 2019.

\bibitem{cyclegan}
J. Zhu, T. Park, P. Isola, and A. Efros. "Unpaired image-to-image translation using cycle-consistent adversarial networks." In Proceedings of the IEEE international conference on computer vision, pp. 2223-2232. 2017.
\bibitem{midinet}
L. Yang, S. Chou, Y. Yang, "MidiNet: A Convolutional Generative Adversarial Network for Symbolic-domain Music Generation," arXiv preprint arXiv:1703.10847, 2017.

\bibitem{seqgan}
L. Yu, W. Zhang, J. Wang, and Y. Yu. "Seqgan: Sequence generative adversarial nets with policy gradient." In Thirty-First AAAI Conference on Artificial Intelligence. 2017.

\bibitem{cgantime}
R. Fu, J. Chen, S. Zeng, Y. Zhuang, and A. Sudjianto. "Time Series Simulation by Conditional Generative Adversarial Net." arXiv preprint arXiv:1904.11419 (2019).

\bibitem{doppelganger}
Z. Lin, A. Jain, C. Wang, G. Fanti, and V. Sekar. "Generating High-fidelity, Synthetic Time Series Datasets with DoppelGANger." arXiv preprint arXiv:1909.13403 (2019).



\bibitem{dcgan}
A. Radford, L. Metz, and S. Chintala, "Unsupervised representation learning with deep convolutional generative adversarial networks," arXiv preprint arXiv:1511.06434, 2015.
\bibitem{psds}
P.Sauer, M. Pai and J. Chow, "Power System and Dynamic Stability", Wilet IEEE Press, New Jersey, 2017. pp. 140-170.

\bibitem{mam}
Formerly TP462. "Identification of electromechanical modes in power systems." IEEE Task Force Report, 2012.

\bibitem{kundur}
P. Kundur, N. Balu, and M. Lauby. Power system stability and control. Vol. 7. New York: McGraw-hill, 1994.


\bibitem{pca_pmu}
L. Xie, Y. Chen and P. R. Kumar, "Dimensionality Reduction of Synchrophasor Data for Early Event Detection Linearized Analysis," IEEE Trans. Power Syst., vol29, no.26, pp. 2784-2794, Nov 2014.

\bibitem{data_science}
L. Cai and Y. Zhu. "The challenges of data quality and data quality assessment in the big data era." Data science journal 14 (2015).

\end{thebibliography}

\end{document}